\documentclass[sigconf]{aamas} 
\usepackage{balance} % for balancing columns on the final page
\usepackage{subcaption}
\usepackage{epstopdf}
\setcopyright{ifaamas}
\acmConference[AAMAS '24]{Proc.\@ of the 23rd International Conference
on Autonomous Agents and Multiagent Systems (AAMAS 2024)}{May 6 -- 10, 2024}
{Auckland, New Zealand}{N.~Alechina, V.~Dignum, M.~Dastani, J.S.~Sichman (eds.)}
\copyrightyear{2024}
\acmYear{2024}
\acmDOI{}
\acmPrice{}
\acmISBN{}

\acmSubmissionID{1004}

\title[Human Goal Recognition]{Human Goal Recognition as Bayesian Inference: Investigating the Impact of Actions, Timing, and Goal Solvability}

\author{Chenyuan Zhang}
\affiliation{
  \institution{The University of Melbourne}
  \city{Melbourne}
  \country{Australia}}
\email{chenyuanz@student.unimelb.edu.au}

\author{Charles Kemp}
\affiliation{
  \institution{The University of Melbourne}
  \city{Melbourne}
  \country{Australia}}
\email{c.kemp@unimelb.edu.au}

\author{Nir Lipovetzky}
\affiliation{
  \institution{The University of Melbourne}
  \city{Melbourne}
  \country{Australia}}
\email{nirlipo@unimelb.edu.au}

\begin{abstract}
Goal recognition is a fundamental cognitive process that enables individuals to infer intentions based on available cues. Current goal recognition algorithms often take only observed actions as input, but here we use a Bayesian framework to explore the role of actions, timing, and goal solvability in goal recognition. We analyze human responses to goal-recognition problems in the Sokoban domain, and find that actions are assigned most importance, but that timing and solvability also influence goal recognition in some cases, especially when actions are uninformative. We leverage these findings to develop a goal recognition model that matches human inferences more closely than do existing algorithms. Our work  provides new insight into human goal recognition and takes a step towards more human-like AI models.
\end{abstract}


\newcommand{\BibTeX}{\rm B\kern-.05em{\sc i\kern-.025em b}\kern-.08em\TeX}

\makeatletter
\gdef\@copyrightpermission{
	\begin{minipage}{0.3\columnwidth}
		\href{https://creativecommons.org/licenses/by/4.0/}{\includegraphics[width=0.90\textwidth]{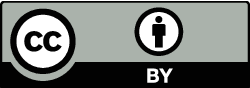}}
	\end{minipage}\hfill
	\begin{minipage}{0.7\columnwidth}
		\href{https://creativecommons.org/licenses/by/4.0/}{This work is licensed under a Creative Commons Attribution International 4.0 License.}
	\end{minipage}
	\vspace{5pt}
}
\makeatother

\begin{document}

%%% The following commands remove the headers in your paper. For final 
%%% papers, these will be inserted during the pagination process.

\pagestyle{fancy}
\fancyhead{}

%%% The next command prints the information defined in the preamble.

\maketitle 

%%%%%%%%%%%%%%%%%%%%%%%%%%%%%%%%%%%%%%%%%%%%%%%%%%%%%%%%%%%%%%%%%%%%%%%%

\section{Introduction}

Imagine that you are a security guard monitoring a camera feed, and you witness a person approaching a locked door. The situation is compatible with two potential goals: entering the conference room behind the locked door or proceeding to the lounge outside the door. If you observe the person pausing for an extended period outside the door, you might infer that they intended to access the locked conference room, a goal that is currently unachievable. However, if the person passes by the door without stopping, you might infer that they have the achievable goal of visiting the lounge.

As this example illustrates, people's ability to infer the intentions of others may be  influenced by factors such as timing information in addition to observed actions \cite{singh2018combining, gates2021rational, Zhang_Kemp_Lipovetzky_2023}. Furthermore, individuals can sometimes infer goals that the actor cannot currently achieve. However, most existing goal recognition  focus on actions alone, neglecting the broader context, and they struggle to handle situations involving unsolvable goals \cite{zhi2020online,ramirez2010probabilistic,keren2015goal, pereira2017landmark, masters2019goal}. In this paper, we draw on behavioral experiments to explore how goal recognition in humans is influenced by three kinds of information: actions, timing, and goal solvability.

Goal recognition is the problem of inferring an actor's real goal given a sequence of observations and a set of possible goals. Two notable approaches that draw on Bayesian inference~\cite{charniak1993bayesian} have emerged in the literature. In 2009, \citet{baker2009action} introduced the inverse planning Bayesian model, aimed at simulating human plan recognition by modeling human Theory of Mind formally as planning. Around the same time, \citet{ramirez2009plan} independently proposed a generative approach that uses planning algorithms over planning models and is
known as plan recognition as planning (PRP). 

Beyond actions alone, a small group of researchers in AI and cognitive science have explored how additional sources of information help to convey  what others are thinking.  \citet{singh2018combining} used gaze data to infer  people's intentions and discovered that gaze can help uncover the hidden goals of players in a board game. \citet{gates2021rational} developed a Bayesian model that explains how people use response times as a cue to preferences in one-shot decision making situations. \citet{Zhang_Kemp_Lipovetzky_2023} generalized the underlying idea and explored how timing information can be used in situations where actors generate rich sequences of actions, not just one-shot decisions. While both \citet{Zhang_Kemp_Lipovetzky_2023} and \citet{berke2023thinking} report that people are sensitive to timing information, there have been no comprehensive attempts to understand the extent to which timing affects human goal inferences.

Beyond actions and timing, the solvability of candidate goals provides a third relevant cue that may influence people's goal inferences. It seems plausible that people tend to assume actors are working towards achievable goals, because actors often have accurate beliefs and are unlikely to waste effort working towards goals that they believe to be unachievable. To the best of our knowledge, however, there has been little work on the impact of solvability in goal-recognition scenarios. Psychological studies of solvability judgments generally focus on tasks like unscrambling anagrams~\cite{topolinski2016can, novick2003nature}, and planning scenarios have received little attention.  %One reason why solvability judgments are directly relevant to planning is that they allow for a sharp contrast between online planning (i.e.\ acting before generating a full solution) and offline planning (i.e. generating a full solution before acting) in human problem solving. If humans follow an offline planning strategy, they should recognize the unsolvability of a goal before carrying out any actions. from the outset.
We therefore consider solvability in addition to actions and timing information, and develop an experiment that aims to understand how these three factors influence goal inference in humans. 

% In Figure~\ref{fig:examples}, we outline how various factors may influence goal inference. Within the map displayed in Figure~\ref{fig:action_map} that aims to test the effect of action on human goal inference, when individuals observe the key step of moving to the right, they tend to favor the red goal, seeing it as the potential path to its achievement. Conversely, if the key step involves moving left, as shown, participants are inclined toward the green goal, irrespective of solvability concerns. Figures~\ref{fig:easy-goal_map_1} and~\ref{fig:easy-goal_map_2} present two identical maps with different green goal positions, with Figure~\ref{fig:easy-goal_map_1} representing a solvable green goal and Figure~\ref{fig:easy-goal_map_2} an unsolvable one. We can observe how the perception of unsolvability may affect an individual's confidence in selecting a goal in these configurations. The map in Figure~\ref{fig:cmp-path_map} aims to investigate the influence of thinking time on human inference The red  goal has a single viable path, while another goal offers two potential paths. If participants observe a prolonged thinking time at location 1 (the key step), followed by the worker moving upwards, it might lead to a shift in their inference towards the green goal. This is because the extended thinking time could suggest that the worker was deliberating on which path is more optimal, contributing to the shift in goal preference. 

\begin{figure*}[t]
\centering
\begin{subfigure}[t]{0.3\textwidth}
  \centering
      \includegraphics[width=\linewidth]{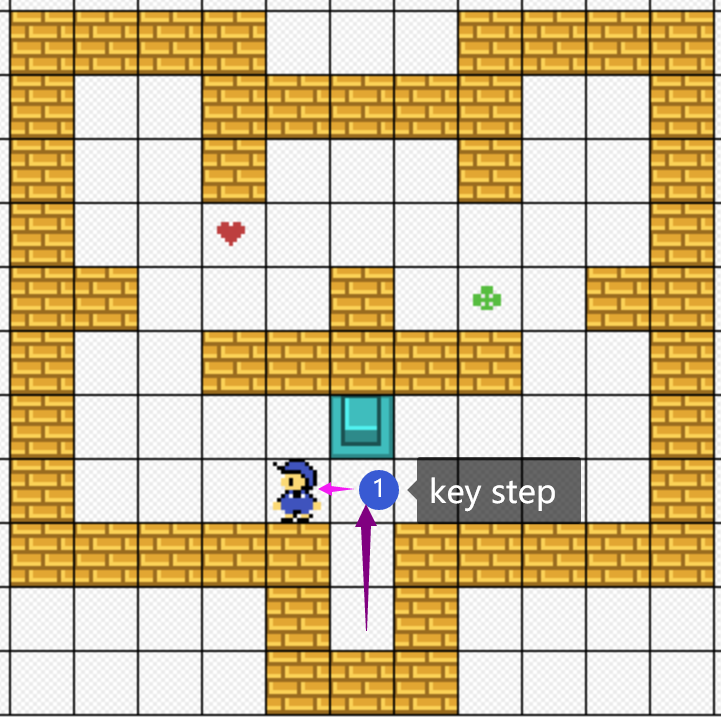}
  \caption{}
  \label{fig:action_map}
 \end{subfigure} 
 \begin{subfigure}[][1pt][t]{0.2\textwidth}
 \vspace{-17.5em}
\centering
  \includegraphics[width=\linewidth]{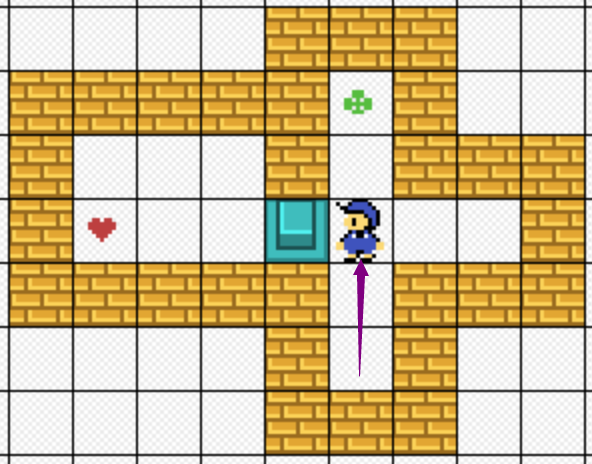}
  \caption{}
  \label{fig:easy-goal_map_2}
\end{subfigure}
\hspace{-0.2\textwidth} % Add some vertical space
\begin{subfigure}[][1pt][t]{0.2\textwidth}
\vspace{-7em}
\centering
  \includegraphics[width=\linewidth]{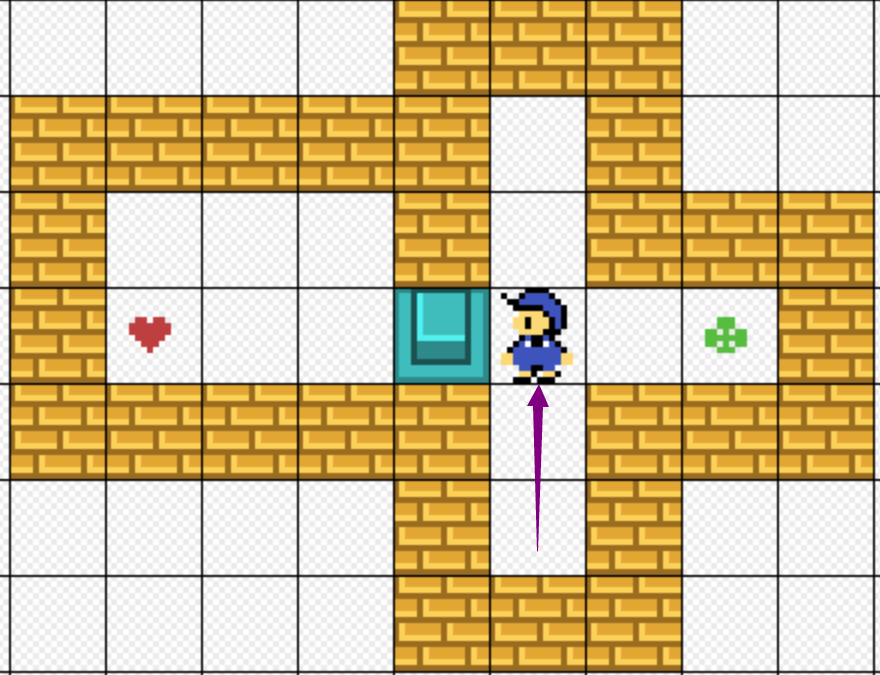}
  \caption{}
  \label{fig:easy-goal_map_1}
\end{subfigure}
\begin{subfigure}[t]{0.4\textwidth}
\centering
  \includegraphics[width=\linewidth]{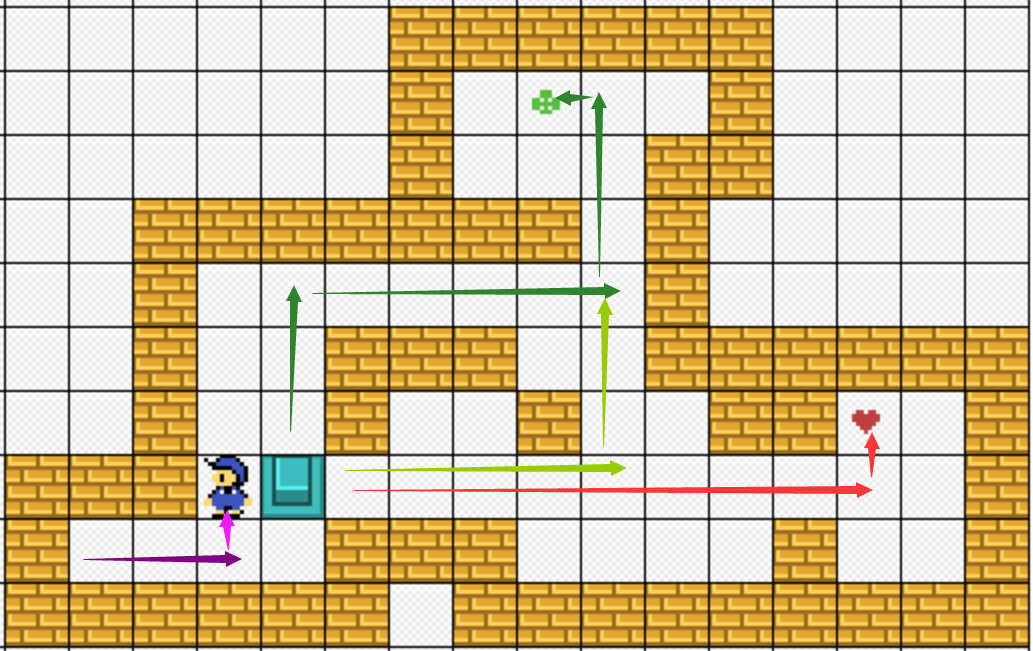}
  \caption{}
  \label{fig:cmp-path_map}
\end{subfigure}
\label{fig:examples}
\caption{Examples showing three types of Sokoban maps in which action, timing and solvability might affect human goal inference. After performing two forced moves indicated by the purple arrows, the actor executes a key step indicated by the pink arrow. Complete sets of maps used in our experiment can be found in the supplementary materials. (a) An \textit{action} map. The red goal is achievable and the green goal is not, and the actor moves left at the key step. (b) An \textit{easy-goal} map. The red goal is easy to achieve but the green goal is not achievable. The key move (not shown) involves a push to the left.  (c) A second \textit{easy-goal} map. The red goal is easy to achieve but the path to the green goal is more complex. At the key move (again not shown) the actor pushes the box to the left. (d) A \textit{competing-path} map. There is one  good path (red arrows) to the red goal and two good paths (green arrows) to the green goal. The actor moves up at the key step.}
\end{figure*}

Figure 1 suggests how the three factors can be studied using goal-recognition tasks within the domain of Sokoban. In all cases the actor is required to push a box towards a goal, and the observer must infer which of two candidate goals the actor is working towards.  Figure~\ref{fig:action_map} is used to study the effect of observed actions. If the actor moves left at the key step shown as a pink arrow, people typically infer that the goal must be the green club, but had the actor moved right instead the red heart would be more probable.  Figures~\ref{fig:easy-goal_map_1} and~\ref{fig:easy-goal_map_2} feature two identical maps with distinct green goal positions, with Figure~\ref{fig:easy-goal_map_1} representing a solvable green goal and Figure~\ref{fig:easy-goal_map_2} an unsolvable one. The probability assigned to the red goal may increase when the green goal is unsolvable rather than solvable.  In Figure~\ref{fig:cmp-path_map}, there is a single potential path towards the red heart but two possible paths towards the green club. If the agent thinks for a long time before taking the key step shown as a pink arrow, one possible inference is that the goal is green and the agent is deciding which of the two paths to pursue. In contrast, the red goal provides no plausible explanation of an extended pause before the key step.

In conventional goal recognition tasks, the evidence typically comprises one or more observed actions. Here, we also consider scenarios where no actions are observed. We refer to these instances as  \textit{prior} instances, because they probe expectations in advance of observing any actions. These prior instances allow us to investigate how solvability influences goal-recognition  when other sources of information  are absent. For instance, in Figure~\ref{fig:action_map}, in the absence of any observations, individuals may exhibit a slight preference for the solvable goal (the red heart). Previous Bayesian models of goal recognition  typically assume a uniform prior~\cite{ramirez2010probabilistic,vered2016online,sohrabi2016plan,baker2009action}, but a small body of recent work has explored how actions observed in previous instances shape the priors that observers apply to new goal-recognition instances~\cite{gusmao2021inferring, masters2021extended}. Here we take a different approach, and explore how the prior reflects structural properties such as solvability and solution complexity rather than previously-observed sequences of actions. 
%In this paper, we introduce a novel prior model that integrates domain-independent structure properties, such as solvability and solution complexity in an attempt to capture the expectations humans bring to goal recognition tasks. In contrast to previous approaches, our approach takes a distinct path, focusing on modeling the non-uniform prior of human inference without relying on historical behavior.

To preview our results, we find that solvability influences people's goal-recognition judgments when no actions have been observed, but that this factor may be subsumed by a more general notion of solution complexity.   When actions are observed, however, solvability appears to play a minimal role, and people's goal-recognition inferences are shaped instead by actions (as a primary factor) and timing information (as a secondary factor). 
We evaluate a suite of formal models and find that human goal inference is well-captured using Bayesian inference, and in particular that a Bayesian model which incorporates an online planner provides a good account of human judgments.

Our work makes several kinds of contributions. First, we carry out a comprehensive behavioral experiment aimed at thoroughly investigating the  factors that influence human goal recognition. This study provides a strong foundation for the development of computational models of human goal inference. Second, we expand upon the planning model introduced by \citet{Zhang_Kemp_Lipovetzky_2023} by integrating a component that allows the planner to recognize unsolvable goals.
%marking an initial effort to replicate human behavior in scenarios involving unsolvable goals within a planning-based framework.
%Third, we demonstrate that human goal inference can be effectively modeled as a Bayesian process. 
Third, we introduce a  human-like goal recognition algorithm that relies on Bayesian inference, and show that it provides a good account of human behavior. 
%that emulates human goal inference behavior, built upon the Bayesian framework.

%%%%%%%%%%%%%%%%%%%%%%%%%%%%%%%%%%%%%%%%%%%%%%%%%%%%%%%%%%%%%%%%%%%%%%%%

% \section{The Preamble}

% You will be assigned a submission number when you register the abstract 
% of your paper on \textit{EasyChair}. Include this number in your 
% document using the `\verb|\acmSubmissionID|' command.

% Then use the familiar commands to specify the title and authors of your
% paper in the preamble of the document. The title should be appropriately 
% capitalised (meaning that every `important' word in the title should 
% start with a capital letter). For the final version of your paper, make 
% sure to specify the affiliation and email address of each author using 
% the appropriate commands. Specify an affiliation and email address 
% separately for each author, even if two authors share the same 
% affiliation. You can specify more than one affiliation for an author by 
% using a separate `\verb|\affiliation|' command for each affiliation.

% Provide a short abstract using the `\texttt{abstract}' environment.
 
% Finally, specify a small number of keywords characterising your work, 
% using the `\verb|\keywords|' command. 

\section{Goal Recognition and Bayesian Framework}
%  Then we introduce a novel planning algorithm that integrates a solvability awareness mechanism, which provides a vital component for the analysis of the human behavior data.

% \subsection{Problem Formulation}

We now formalize the problem of goal recognition and introduce a Bayesian framework for this problem.  We follow the notation commonly used in the planning community \cite{ramirez2009plan, vered2016online, sohrabi2016plan}, but the same general approach has been applied in the cognitive science literature \cite{baker2009action}. Because we consider timing information, our problem formulation includes this information.

\begin{definition}
A goal recognition problem with timing information is a tuple $\langle  D, G, Prior, O \rangle$, where $D$ is a planning domain, $G = \{g_1, g_2, ..., g_n\}$ is a set of possible goals for the planning domain, $Prior$ is the prior probability $P(G)$ over $G$, and $O$ is a sequence of observations $\langle  a_0, t_0\rangle, ..., \langle  a_m, t_m\rangle$, where $a_i\in A$ is an action, and $t_i$ is a non-negative real number denoting the planning time used to select $a_i$ for execution. 
\end{definition}

Goal recognition can then be carried out using
\begin{equation}\label{eq:1}
        P(G|O) \propto Prior(G)LL(O,G),
        \end{equation}
where $P(G|O)$ is the posterior distribution over goals,  $Prior(G)$ is the prior $P(G)$ and $LL(O,G)$ the likelihood 
$P(O|G)$. Following \cite{Zhang_Kemp_Lipovetzky_2023}, we decompose the likelihood $LL(O,G)$ into two components: the timing component $LL_T(O,G)\coloneqq P(\langle t_0, t_1, ..., t_m\rangle|G)$ and the action component $LL_A (O,G)\coloneqq P(\langle a_0, a_1, ..., a_m\rangle|G)$, allowing for independent calculations.

While solvability, actions, and timing might all influence human goal inference, a Bayesian perspective suggests a fundamental distinction between solvability and the other two factors.   Solvability is an inherent property of the goal and should therefore be captured by $Prior(G)$ within the Bayesian model. In contrast, actions and timing are aspects of $O$, the observation sequence, and should be incorporated in the likelihood  $LL(O,G)$. 

Because previous Bayesian accounts of goal recognition usually assume a uniform prior%$Prior_\mathcal{U}(\cdot)=\frac{1}{|G|}$ 
~\cite{ramirez2010probabilistic,vered2016online,sohrabi2016plan,baker2009action}, they focus on estimating the likelihood term $LL(O,G)$. Specifically, this involves determining the probability of generating the provided observation sequence $O$ given the goal $G$. Most goal recognition models rely on standard planning algorithms that do not handle scenarios in which the goal $G$ is unsolvable. For example, with classical planning, unsolvable goals are filtered out from consideration at the outset. 
 
 Some approaches avoid the assumption that the actor is rational \cite{vered2016online, Zhang_Kemp_Lipovetzky_2023, zhi2020online, baker2009action}, and can therefore estimate the likelihood of an unachievable goal. We go beyond these approaches by using a novel solvability-aware planner (i.e. solvability-aware Adaptive Lookahead Planner, full details in the \href{https://drive.google.com/file/d/1I4Yb2luisemuycio2REInIqX80can5X8/view?usp=drive_link}{supplementary information}) that can decide whether a goal is unsolvable.
 %a mechanism to determine when to terminate the search process if they have gather enough evidence, similar to how humans make such decisions. Therefore, in this paper, we present a novel solvability-aware planning algorithm (refer to the supplementary material for the complete pseudocode), which builds upon the Adaptive Lookahead Planner\cite{zhang2023comparing} to address this challenge. 
 While searching through the state space, the planner maintains a closed list of previously visited states in memory, and terminates and declares the goal unachievable if no new states are encountered during a fixed number of iterations. 
 %This model is a conceptual framework designed to potentially capture human responses towards unsolvable goals. 
 Although we provide a limited evaluation of this planner as an account of human planning, our primary focus is on evaluating the Bayesian model of goal recognition that incorporates this planner as a likelihood estimation component.  %While we do provide a comparison between this model and human behavior in the subsequent section, the development of a comprehensive solvability awareness framework for human-like behavior is itself a complex task and it is not the central focus of this study. We employ this model exclusively as a tool to estimate likelihoods within the Bayesian framework of the goal recognition models. 

%%%%%%%%%%%%%%%%%%%%%%%%%%%%%%%%%%%%%%%%%%%%%%%%%%%%%%%%%%%%%%%%%%%%%%%%

\section{Experiment Configuration}

To explore how actions, timing and solvability influence goal recognition and to test competing computational models we conducted a human experiment using the Sokoban domain. Although goal recognition is our primary focus, the experiment began with a planning phase in which participants were asked to solve 23 Sokoban problems. 9 of these problems were unsolvable, and participants could press a specified button at any stage if they believed that the current instance was unsolvable. 

Participants then moved on to a goal-recognition phase using the same maps presented in the planning phase. 
%involving problem-solving and goal recognition within the Sokoban domain. 
% In this context, we formulated three qualitative hypotheses: firstly, individuals tend to favor solvable goals, applying solvability as an initial cue before any actions are observed; secondly, observed actions can significantly influence goal inference, even when some of the goals are unsolvable; and thirdly, observed thinking time plays a role in human goal inference, though to a lesser extent than observed actions. 
%All participants responded to the same goal recognition instances, which was presented in a random order. Prior to encountering these instances, participants were engaged in a problem-solving phase involving Sokoban puzzles on the same maps. In the problem-solving phase, there were 23 instances, of which 9 were unsolvable. Participants are consistently given the option to press a button if they believe the instance is unsolvable. This means that in instances that are actually solvable, they might choose to give up. In the goal recognition phase, participants began by pressing a button to start the instance. 
Each instance presented a Sokoban map with two possible goal positions marked as A and B. Participants were asked to infer the actor's goal, and provided responses on a six point Likert scale labeled ``very confident B'', ``fairly confident B``, ``slightly confident B,`` ``slightly confident A,'' and so on.
%After observing the worker's behavior (if present), participants needed to choose their inference along with a confidence level from six options: very confident A, fairly confident A, slightly confident A, slightly confident B, fairly confident B, and very confident B. 
%This was represented on a 6-point Likert scale,
%
For subsequent analyses we mapped these six responses to probabilities \{0, 0.2, 0.4, 0.6, 0.8, 1\}, where each probability represents the probability of choosing goal A \cite{harpe2015analyze}. %In our subsequent analysis, we use probability values choosing goal A in the set {0, 0.2, 0.4, 0.6, 0.8, 1} to simplify the representation of confidence levels. 
For example, if three participants chose ``very confident A'' and two chose ``slightly confident B'' the average response would be $\frac{3\times1 + 2\times0.4}{2+3}=\frac{3.8}{5}=0.76$.
%threethe number of responses is (2, 2, 5, 6, 10, 11) respectively for an instance, then the average confidence level is $\frac{2\times0 + 2\times0.2+5\times0.4+6\times0.6+10\times0.8+1\times11}{2+2+5+6+10+11}=\frac{25}{36}=0.69$.

The stimuli for the goal recognition phase belong to one of three types, and included 20 \textit{prior} instances, 40 \textit{observation} instances and 9 \textit{filler} instances. The presentation order of these instances was fully randomized.
Identical map configurations and goal positions were used for the prior and observation instances, but the prior instances required participants to infer the actor's goal without having observed any actions.  In all \textit{filler} instances participants observed the player pressing the button to declare the instance unsolvable. Responses to these instances will not be analyzed, and they were included only to reinforce the possibility that the goal might be unsolvable.

The \textit{observation} instances included pairs that share identical maps and potential goal positions but differ in a single key step (see Figures~\ref{fig:action_map} and \ref{fig:cmp-path_map}). This key step refers to the first step at which a player who does not backtrack has multiple options.
%including the informative action or the allocation of planning time for goal inference. 
Within each pair, either the action for this step or the response time for the action at this step can vary. There are 20 pairs in total, corresponding to the 20 instances in the \textit{prior} type. 

The observation instances can be organized into three subtypes. 
%There are three specific subtypes within this category. We focused on examining the impact of action in the first subtype \textit{action} pairs, while the effect of timing was explored in the remaining two subtypes. 
\textit{Action} pairs differ based on the action taken  for the key step (see Figure~\ref{fig:action_map}). We hypothesize that changing the action at this step will influence human inferences regardless of the solvability of the potential goals. 

The remaining two subtypes allow us to study the influence of timing information.
\textit{Easy-goal} pairs use maps where one goal is easy to solve and the other goal is either solvable or unsolvable (Figure~\ref{fig:easy-goal_map_1} and \ref{fig:easy-goal_map_2}). In this subtype, the thinking time for the key step varies. We hypothesize that increasing the thinking time at this step will decrease the participant's confidence that the actor is aiming for the easy goal, because achieving the easy goal should not require a prolonged pause at any stage.
%This subtype aims to examine how thinking time can influence human inference when one of the goals is easy to achieve. %Specifically, if participants observe the worker pause for an extended period at the indicated position before pushing the box in the left direction, it might cause a shift in their inference from the red goal to the green goal as the red goal does not need a prolonged thinking time. 
%In summary, our hypothesis suggests that prolonged thinking time will decrease human preference for the easy goal. 

\textit{Competing-path} pairs (the third and final subtype) include cases in which one goal (e.g.\ the green goal in Figure~\ref{fig:cmp-path_map}) requires a choice between two possible actions at the key step, but the other goal suggests only one natural action at this step. As for easy-goal pairs, we vary the thinking time observed at the key step. We hypothesize that increasing the thinking time at this step will suggest that the actor is choosing between two paths, and therefore aiming for the competing-path goal rather than the alternative.
%If participants observe a prolonged thinking time at location 1 (the key step), followed by the worker moving upwards, it might lead to a shift in their inference towards the green goal. This is because the extended thinking time could suggest that the worker was deliberating on which path is more optimal, contributing to the shift in goal preference. Overall, we propose that prolonged thinking time will decrease human preference for the goal with a single option.

For each map configuration, we started with a goal-recognition instance featuring two solvable goals. We then created additional instances by moving each solvable goal in turn to either an adjacent unsolvable position or an unsolvable position with similar properties (e.g.\ Manhattan Distance from the start position). Figure~\ref{fig:easy-goal_map_1} shows an original instance with two solvable goals, and Figure~\ref{fig:easy-goal_map_2} is a variant in which the green goal is unsolvable. Manipulating solvability in this way allows us to explore the influence of solvability on human goal inference.

The experiment was pre-registered on 
\href{https://aspredicted.org/zi55w.pdf}{AsPredicted}. We recruited 100 standard sample participants (63 females and 37 males with a median age of 28) on Prolific, and 5 were excluded because they had more than 3 abnormal responses in the problem solving phase. For each instance, responses more than 3 standard deviations away from the mean total time and total steps for that instance were considered abnormal.

\section{Human Problem Solving Behaviour}
The problem-solving phase in the experiment serves three primary purposes. Firstly, it aims to validate the effectiveness of our manipulation of observations (i.e. actions or thinking times) in the goal recognition phase. Secondly, it seeks to analyze participants' strategies when faced with an unsolvable goal. Lastly, it involves a comparative assessment of the performance between solvability-aware Adaptive Lookahead Planner (A-LH) and human participants across Sokoban instances. The number of iterations generated by A-LH was converted into seconds by normalizing, ensuring that the total planning time for all instances remains the same across humans and A-LH.

% To begin, we examine participants' strategies when faced with solvable instances. We chart the percentage of optimal solutions, along with the average thinking time for the initial three steps, as depicted in Figure \ref{fig:ps_sol}. Notably, instances situated in the right-bottom corner exhibit a trend of being relatively easy for participants. In addition to the evident pattern indicating that easy goals within easy-goal maps prove to be particularly simple, there is a broader observation: the top region presents a somewhat more challenging aspect for participants compared to the right-sided area in the competing-path maps. In the case of unsolvable instances, there doesn't appear to be any discernible pattern based on the map type.

% \begin{figure}[t]
% \centering
% \begin{subfigure}[t]{0.4\textwidth}
% \centering
%   \includegraphics[width=\linewidth]{AnonymousSubmission/LaTeX/ps_sol.png}
%   \caption{}
%   \label{fig:ps_sol}
% \end{subfigure}
% \hfill
% \begin{subfigure}[t]{0.4\textwidth}
% \centering
%   \includegraphics[width=\linewidth]{AnonymousSubmission/LaTeX/ps_unsol.png}
%   \caption{}
%   \label{fig:ps_unsol}
% \end{subfigure}
% \caption{(a) Overview of participant's performance on solvable instances  (b) Overview of participant's performance on solvable instances}
% \label{fig:action}
% \end{figure}

\begin{figure*}[t]
\centering
\begin{subfigure}[t]{0.2\textwidth}
\centering
  \includegraphics{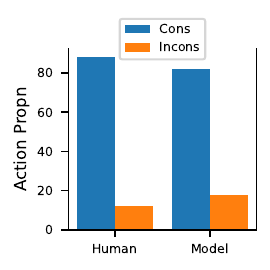}
  \caption{}
  \label{fig:ps_action}
\end{subfigure}
\hfill
\begin{subfigure}[t]{0.24\textwidth}
\centering
  \includegraphics{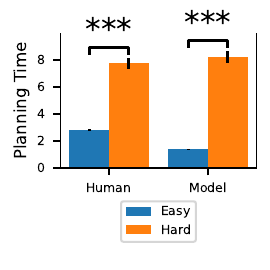}
  \caption{}
  \label{fig:ps_easy}
\end{subfigure}
\begin{subfigure}[t]{0.24\textwidth}
\centering
  \includegraphics{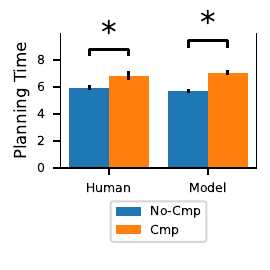}
  \caption{}
  \label{fig:ps_cmp}
\end{subfigure}
\begin{subfigure}[t]{0.24\textwidth}
\centering
  \includegraphics[width=\linewidth]{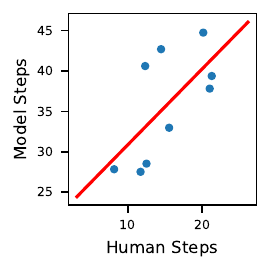}
  \caption{}
  \label{fig:ps_step}
\end{subfigure}
\caption{Results for the planning phase. (a) Proportion of participant choices for the action in \textit{action} maps. \textit{Cons} means consistent with our manipulation in the goal recognition phase. The model employs softmax action selection with a temperature parameter set to 5. (b) Average Planning time for \textit{easy} and \textit{hard} goals in \textit{easy-goal} maps. The effect of thinking time is significant for both humans and the model ($p<0.001$). For both (b) and (c), error bars show the standard deviation of the mean planning time (measured in seconds). (c) Average Planning time for \textit{competing} and \textit{no-competing} goals in \textit{competing-path} maps. The effect of thinking time is significant for both human and model ($p<0.05$). (d) Number of steps taken in unsolvable instances for humans (x-axis) and the model (y-axis). Human responses and model predictions are strongly correlated ($r(7)=0.65, p=0.05$).}
\label{fig:ps_subtype}
\end{figure*}

Across the \textit{action} maps, the majority of participants (88\%) make choices that match our manipulation in the goal recognition phase, which is consistent with the model's prediction (82\%) as shown in Figure~\ref{fig:ps_action}. Across  \textit{easy-goal} maps (e.g.\  Figure~\ref{fig:ps_easy}), participants spend less time on the easy goal, with an average of 2.84 seconds compared to 7.75 seconds for the harder goal. The model's prediction shows a similar trend: 1.41 seconds for the easy goal and 8.24 seconds for the hard goal. Across \textit{competing-path} maps, both human participants and the model show a small but statistically significant difference in planning times for the two goals. Human planning times increase from 5.94 seconds to 6.83 seconds, and the model predicts an increase from 5.68 to 7.04 seconds (see Figure~\ref{fig:ps_cmp}). These results indicate that the  manipulations in our experiments are well-grounded and also suggest that the A-LH planner provides a good account of human behavior in the Sokoban domain.

We further examined the number of steps taken before participants became aware that unsolvable instances were in fact unsolvable. The results in Figure~\ref{fig:ps_step} demonstrate a positive correlation between the model's predictions and human responses. The majority of participants demonstrated behavior resembling that of online planners, taking an average of 15.23 steps, indicating that participants take approximately 15 steps before recognizing the unsolvability of the goal. The model predicted a much higher average of 35.78 steps. This divergence might be attributed to participants' general lack of patience during online experiments. A minority of participants do recognize goals as unsolvable before carrying out any actions, and failing to capture the responses of these participants may also contribute to the difference between model predictions and average human responses.
%appear to resemble offline planners, and failing to The model's oversight in accounting for some participants as offline planners (i.e.\ realizing unsolvability at the outset) might also contribute to this difference. 
Nevertheless, our data strongly suggest that the majority of people should be characterized as online planners in our experimental context.

\section{Human Goal Recognition}
We use mixed effects models to fit the human responses in the goal recognition phase. In these models, the variable \textit{CL} represents the confidence level towards goal A, ranging from 0 to 1. The variables \textit{soA} and \textit{soB} correspond to the solvability of goals A and B, respectively, with 1 denoting solvability and -1 denoting unsolvability. In the \textit{action} maps, goal A represents the rightmost goal, while goal B represents the leftmost goal.
%with an almost symmetric arrangement. 
In \textit{easy-goal} maps, goal A is designated as the easy goal, while goal B is identified as the hard goal. In \textit{competing-path} maps, goal A signifies the no-competion goal, while goal B denotes the competing-paths goal. The variable \textit{obs} indicates whether the observation (i.e. action or planning time) is consistent with goal A (1 denotes consistent, -1 denotes inconsistent) if available. The model also includes random effects for participant and map configuration. All p-values subsequently reported are based on log-likelihood ratio tests, where $M_0$ serves as the null model.  The models and summary of regression results can be found in Table~\ref{table1}. 

\subsection{Prior Instances}
\begin{figure}[t]
\begin{subfigure}[t]{0.23\textwidth}
\centering
  \includegraphics[width=\textwidth]{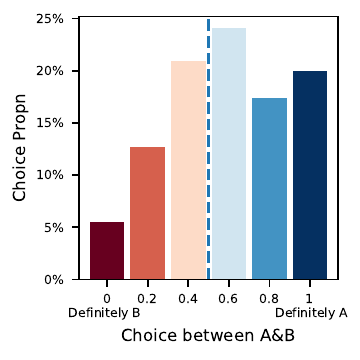}
  \caption{}
  \label{fig:prior_overall}
\end{subfigure}
\begin{subfigure}[t]{0.23\textwidth}
\centering
  \includegraphics{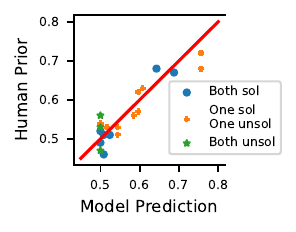}
  \caption{}
  \label{fig:model_prior}
\end{subfigure}

\begin{subfigure}[t]{0.47\textwidth}
\centering
  \includegraphics{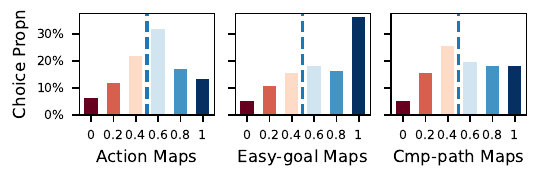}
  \caption{}
  \label{fig:prior_subtype}
\end{subfigure}
\caption{ Results for prior instances in the goal-recognition phase. (a) Response distribution for prior instances where goal A is solvable and goal B is not. Blue bars indicate a preference for solvable goal A while red bars represent a preference for unsolvable goal B. (b) Comparison between human responses and the easiness model. The x-axis represents the model's predicted probability of choosing the easy goal, and the y-axis represents the human prior observed in the experiment. The instances are represented as circles, crosses or stars based on whether neither, one or both goals are unsolvable. (c) Response distribution from  panel (a)  broken down by the three subtypes.}
\label{fig:prior}
\end{figure}

In \textit{prior} instances, we present a map without any observed actions to determine how solvability or other static properties would influence the human prior $Prior_\mathcal{H}(G)$ over the potential goals. Our hypothesis is that humans will prefer solvable goals in  cases where one goal is solvable and the other is unsolvable.
As shown in Figure~\ref{fig:prior}, the overall choice percentage of solvable goals stands at 61.16\% (the sum of blue bars), and the average response is 0.59. This result confirms a clear preference for goals that can be solved.

\begin{table*}[t]
\centering
\begin{tabular}{l|c|c|c|c|c}
\hline
Model  & Model String ($CL \sim$)& Prior & Action & Easy-goal & Competing-path \\
\hline
M0 & $(1|participant) + (1|map) $ & 6762.8 & 6252.6 & 2591.2 & 5463.8 \\
M1 & $soA + soB + soA*soB + (1|participant) + (1|map) $ & \textbf{6741.3} & 6272.5 & 2597.1 & \textbf{5439.5} \\
M2 & $obs + (1|participant) + (1|map) $ & N/A & \textbf{4621.3} & \textbf{2552.4} & 5466.2 \\ 
M3 & $soA + soB + soA*soB + obs + (1|participant) + (1|map) $ & N/A & 4636.9 & 2558.3 & 5441.7 \\
\end{tabular}
\caption{Bayesian Information Criterion (BIC) scores of models in regression analysis. The best model for each set of instances (i.e.\ each column) is shown using  bold. The dependent variable $CL$ is the probability assigned to goal A.}
\label{table1}
\end{table*}

% \begin{figure}[t]
%     \centering
%   \includegraphics{AAMAS-2024-Formatting-Instructions-CCBY/model_prior_new.png}
%   \caption{Comparison of Easiness Prior Model and Human Prior. The x-axis represents the model's predicted probability of choosing the easy goal, and the y-axis represents the human prior observed in the experiment. The instances are represented as circles, crosses, or stars based on whether both goals are solvable, one is solvable while the other is unsolvable, or both are unsolvable, respectively.}
%   \label{fig:model_prior}
% \end{figure}

 The log-likelihood ratio test of prior instances yields $\chi^2(3) = 44.185, p<0.001$. Model M1 demonstrates a strong fit, implying that the impact of solvability is evident. Specifically, the 95\% Confidence Interval (CI) for the regression coefficient of soA is [-0.05, -0.02], while the 95\% CI for soB is [0.02, 0.05]. These findings confirm our hypothesis — when one target is solvable, participants are more likely to infer that the solvable target represents the actual goal.

When we look deeper into the differences between various types of scenarios, we notice that distinct map layouts affect how much participants rely on  solvability (see Figure~\ref{fig:prior_subtype}). Specifically, in the \textit{action} maps, where the primary contrast between the goals is solvability, a consistent pattern emerges: participants tend to lean toward solvable goals. Most participants, however,  express only a slightly confident viewpoint. This suggests that even though participants recognize the importance of solvability, the evidence supporting it might not be strong enough to firmly guide their conclusions.

In the \textit{easy-goal} maps, the findings reveal a substantial number of participants who exhibit strong confidence in favor of the target being solvable rather than unsolvable. This finding, however, prompts the question of whether this confidence stems solely from solvability or is influenced by other characteristics within the \textit{easy-goal} maps. As mentioned already, within these maps the solvable goal coincides with the easier goal. In order to further explore the possible role of easiness, we compared responses to maps that were similar except that the hard goal was solvable rather than unsolvable. We found that solvability itself does not significantly impact human inference; rather, individuals consistently lean towards the easier goal, irrespective of the solvability status of the other goal. 

For \textit{competing-path} maps,  solvability continues to shape human judgments, but in a different way. Among the responses, 54.74\% show a preference for the solvable goal, resulting in a mean confidence level of 0.57. This is even higher than the 0.56 confidence level in the \textit{action} maps. Interestingly, when participants choose a solvable goal, their behavior stands out from when they pick an unsolvable one. While they don't seem very sure about choosing an unsolvable goal, their confidence is more balanced when they opt for a solvable goal.

Our findings indicate that human goal inference is shaped by both the solvability and inherent difficulty of goals. An unsolvable goal might represent an extreme version of a difficult goal. To test this idea, we developed a simple  model called the \textit{Easiness Prior Model} to fit the human prior. In this model, we operationalize the difficulty of each solvable goal $g$ as the sum of the optimal path length $opt(g)$ and a smoothing parameter $o$ (set to 5 in our analyses). This parameter captures the baseline cognitive effort demanded by the task (e.g.\ effort to process the map, recognize the actor and goal locations, etc). We further assume that unsolvable goals have the same difficulty score ($c=26$) as the most difficult solvable goal in our experiments. 
%Hence, we assign the largest difficulty score ($s=26$ in our experiment) among solvable goals.
Overall, the difficulty score for goal $g$ is defined as $s_g = o + min(c, opt(g))$. Let $s_A$ and $s_B$ represent the cognitive difficulty values for goals A and B respectively in the prior instances. To reflect the notion that easier goals (with shorter optimal paths) have a higher prior, we use 
\vspace{-0.5em}
\begin{equation}\label{eq2}
    \langle Prior(A), Prior(B) \rangle= \langle \frac{s_B}{(s_A+s_B)},\frac{s_A}{(s_A+s_B)} \rangle.
\end{equation} As shown in Figure~\ref{fig:model_prior}, our model closely aligns with the actual prior probabilities observed in the prior instances (Pearson correlation test: $r(18)=0.91, p<0.001$). This finding suggests that our simple easiness model can effectively mimic human decision-making when no observations are available. %%% More explaination?

\subsection{Observation Instances}
The observation instances consists of pairs that share identical maps and potential goal configurations but differ in a single key step. This key step refers to the first action where a player who does not backtrack has multiple options. Within each pair, either the action for this step or the response time for the action can vary. Each pair also corresponds to a prior instance which shares the same map and goal configurations without including any observations.

There are three specific subtypes within the observation pairs, which also corresponds to three different types of maps in the prior instances. In what follows we consider the three subtypes separately.  

\subsubsection{Action Pairs}

% \begin{figure*}[t]
% \centering
% \begin{subfigure}[t]{0.33\textwidth}
% \centering
%   \includegraphics[width=1.1\linewidth]{AAMAS-2024-Formatting-Instructions-CCBY/action_new.png.png}
%   \caption{}
%   \label{fig:action}
% \end{subfigure}
% \hfill
% \begin{subfigure}[t]{0.33\textwidth}
% \centering
%   \includegraphics[width=1.1\linewidth]{AAMAS-2024-Formatting-Instructions-CCBY/easy_new.png}
%   \caption{}
%   \label{fig:easy}
% \end{subfigure}
% \hfill
% \begin{subfigure}[t]{0.33\textwidth}
% \centering
%   \includegraphics[width=1.1\linewidth]{AAMAS-2024-Formatting-Instructions-CCBY/cmp_new.png}
%   \caption{}
%   \label{fig:cmp}
% \end{subfigure}
% \caption{(a) Percentage of participant choice of \textit{action} pairs.  (b) Percentage of participant choice of \textit{easy-goal} pairs. (c) Percentage of participant choice of \textit{competing-path} paris}
% \label{fig:participant}
% \end{figure*}

% As shown in Figure~\ref{fig:action}, this result confirm our impression: solvability rarely contribute to the final decision in goal choice when action are informative. Regardless of whether the goal is solvable or unsolvable, the shift in goal preference, compared to the prior (that slightly prefer the solvable goal), aligns with the guidance provided by action observations. When the action moves to the unsolvable goal, the confidence level to the solvable goal shift from 0.56 to 0.24, and when the action moves to the solvable goal, the confidence level to that goal increase to 0.81. We also run the log-likelihood ratio test to verify the hypothesis (see Table~\ref{table1}).

The result confirm our hypothesis: solvability rarely contributes to the final decision in goal choice when actions are informative. Regardless of whether the goal is solvable or unsolvable, the shift in goal preference, compared to the prior (that slightly favors the solvable goal), aligns with the guidance provided by action observations. When the action moves to the unsolvable goal, the confidence level to the solvable goal shifts from 0.56 to 0.24, and when the action moves to the solvable goal, the confidence level to that goal increases to 0.81. We also ran a log-likelihood ratio test to verify the hypothesis (see Table~\ref{table1}).

Among the models considered, Model M2 demonstrates the best fit (($\chi^2(1) = 1638.7, p<0.001$)), as evidenced by its lowest Bayesian Information Criterion (BIC) value. The 95\% CI for the regression coefficient of \textit{obs} falls within the range of [-0.32, -0.3]. Conversely, neither \textit{soA} nor \textit{soB} contributes meaningful information to the confidence level in this context. Notably, Model M1 even exhibits a higher BIC value than the baseline model (M0), indicating that solvability fails to enhance the model fit.

% It is important to highlight that quantifying the \textit{informativeness} of action observations in this context from conventional practices within the planning community. In approaches such as PRP and subsequent methods, actions cannot be employed to identify unsolvable goals. 

\subsubsection{Easy-goal Pairs}
% In \textit{easy-goal} instance, one goal is easy to solve and the other goal is either solvable or unsolvable. In this subtype, the thinking time of the key step varies. Our hypothesis suggests that prolonged thinking time will decrease human preference for the easy goal. 

% \begin{figure}[t]
%     \centering
%   \includegraphics[width=\linewidth]{AnonymousSubmission/LaTeX/obvious_overall.png}
%   \caption{Percentage of participant choice of \textit{obvious} pairs}
%   \label{fig:obvious_overall}
% \end{figure}

Compared to the prior condition, regardless of the time actors take to think about the key steps, human responses shift towards the easy goal in the presence of observations. This shift is evident as the confidence level to the easy goal changes from 0.69 to 0.86 given short thinking time. However, when a long thinking time (consistent with hard goals in our hypothesis) is observed, this shift is somewhat less pronounced (0.69 to 0.75). Additionally, we observed that this pattern remains consistent, irrespective of whether the hard goal is solvable or not.

We performed an identical log-likelihood ratio test using easiness to define \textit{obs}, where short thinking time is consistent with easy goal A (assigned 1) and long thinking time is aligned with hard goal B (assigned -1). The results aligned with our initial intuition: Model M2 exhibits the most favorable fit ($\chi^2(1) = 45.425, p<0.001$). This underscores the notion that thinking time is relevant for predicting the confidence level, while solvability's contribution remains negligible. The 95\% confidence interval for the regression coefficient of the intercept spans from 0.26 to 0.34, indicating a strong tendency among participants to favor the easier target choice. Furthermore, the 95\% confidence interval for the regression coefficient of \textit{obs} (i.e. long/short thinking time) falls within the range of [0.04, 0.07]. This outcome emphasizes that the manipulation of thinking time can exert a notable influence on the confidence level, contributing to statistically significant variations in participants' goal inference processes.

\subsubsection{Competing-path Pairs}
% In \textit{Competing} pairs, one goal has an obvious option in the key step, while the other goal presents two competing options. Similar to the \textit{easy-goal} pairs, the thinking time of the key step varies. We propose that prolonged thinking time will decrease human preference for the goal with a single option.

% TODO: a plot with cmp-path and action to show the solvability affect people's choice

% TODO: caption and integration with other two subtypes
% \begin{figure}[t]
% \centering
% \begin{subfigure}[t]{0.4\textwidth}
% \centering
%   \includegraphics[width=\linewidth]{AnonymousSubmission/LaTeX/competing_overall.png}
%   \caption{}
%   \label{fig:cmp_overall}
% \end{subfigure}
% \hfill
% \begin{subfigure}[t]{0.4\textwidth}
% \centering
%   \includegraphics[width=\linewidth]{AnonymousSubmission/LaTeX/competing_long_short.png}
%   \caption{}
%   \label{fig:cmp_sol}
% \end{subfigure}
% \begin{subfigure}[t]{0.4\textwidth}
% \centering
%   \includegraphics[width=\linewidth]{AnonymousSubmission/LaTeX/competing_sol_unsol.png}
%   \caption{}
%   \label{fig:cmp_sol}
% \end{subfigure}
% \caption{(a) Percentage of participant choice of \textit{competing} pairs, A refers to the target in the left area while B refers to the target in the right area  
% (b) Percentage of participant choice of \textit{competing} paris based on thinking time (c) Percentage of participant choice on solvability}
% \label{fig:cmp}
% \end{figure}

Broadly speaking, the patterns observed within the \textit{competing-path} pairs align closely with those of the \textit{easy-goal} pairs. In particular, when participants observe the actions, their preferences shift towards the no-competing goals whether the actor spend more or less time. Unlike the \textit{easy-goal} instances, the initial distribution of \textit{competing-path} maps is nearly uniform (with confidence level to the no-competition goal of 0.5) as shown in Figure~\ref{fig:prior_subtype}. With consistent observations (i.e. short thinking time) favoring the no-competition goal, the confidence level to that goal increases to 0.58. Surprisingly, even with inconsistent observations (i.e. long thinking time), the confidence level still increases to 0.56. This result implies that our definition of consistency (long/short thinking time) may not be the primary factor observers take into account during goal inference. 

%Rather, it can potentially be attributed to the strong sampling assumption \cite{navarro2012sampling}: when individuals observe a path leading towards a goal with that path only, they tend to perceive the goal as more likely compared to goals with multiple available paths.

% To explore whether the solvability factor influences inferences in this scenario (see Figure~\ref{fig:cmp_sol}), we find a slight preference towards solvable goals, closely resembling the results in Figure~\ref{fig:prior_subtype}. These findings suggest that observation does not significantly augment human inferences when the solvability are different between two goals, in contrast to the prior setup.

We applied the same log-likelihood test using number of competing path as the standard to define \textit{obs}, where short thinking time is consistent with no-competing goal A (assigned 1) and long thinking time is aligned with competing-path goal B (assigned -1). All three models yield significantly better fits than the baseline model. Model M1, which considers only solvability, achieves the optimal fit ($\chi^2(2) = 41.442, p<0.001$) based on BIC scores. In the comprehensive Model M3, the 95\% confidence interval for the regression coefficient of \textit{soB} lies between [-0.07, -0.04], while the intervals for \textit{soA} and \textit{obs} are [0.00, 0.04] and [0.00, 0.03] respectively. These results indicate that in this context, the solvability of competing goal B presents a substantial impact on human inferences, while the solvability of the no-competing goal A and the influence of thinking time are comparatively more modest. Increased awareness of solvability of competing goals suggests individuals may allocate more time to plan for these goals, aligning with the assumption in A-LH~\cite{zhang2023comparing, Zhang_Kemp_Lipovetzky_2023}.

\subsection{Model Comparison}
We now evaluate a range of models by comparing them against human goal inference behavior. These models are formulated within the same Bayesian framework (Equation~\ref{eq:1}) but use 3 different priors $Prior(\cdot)$ (uniform prior, easiness prior model shown in Equation~\ref{eq2}, and empirical prior from our problem solving data) and 5 different likelihoods $LL(\cdot,\cdot)$ (offline-planning likelihood, online-planning likelihood, online-planning likelihood with actions only, empirical likelihood, and empirical likelihood with actions only).

For offline-planning likelihood estimation, we adopt the PRP approach outlined by \citet{ramirez2010probabilistic}. This approach is not designed to handle unsolvable goals, 
%and  assigns an In scenarios where a g unsolvable, the original formulation does not account for this, resulting in an assigned infinite cost difference. This ensures that the 
but as originally formulated it consistently prioritizes solvable goals ahead of unsolvable goals. All easy-goal and competing-path maps were intentionally designed so that actions would be uninformative about the goal, and in these cases the offline likelihood assigns equal weight to both targets.

The online likelihood is derived from 100 simulations conducted using the A-LH \cite{Zhang_Kemp_Lipovetzky_2023}. To minimize variance in our model predictions, we focus solely on the likelihood associated with the key step, as the other two steps are predetermined. Specifically, we need to calculate the action component $LL_A$ and the timing component $LL_T$ separately for each goal and then combine them. For the action component, the likelihood is estimated by dividing the number of action choices made in the simulations by the total number of simulations (i.e. 100). As illustrated by Figure~\ref{fig:ps_action}, we previously confirmed that A-LH aligns with human action choices in \textit{action} instances. In the remaining two types of instances, we found that actions still provide valuable information for goal inference. 
%Regarding the estimation of the timing component, a challenge arises in aligning the model's predicted timing, measured in number of iterations, with real-world timing, measured in seconds, when predicting the likelihood for planning time. 
In \textit{action} instances, since the simulated times for both targets are the same, the timing likelihood $LL_T$ effectively makes no contribution. In \textit{easy-goal} and \textit{competing-path} instances, the timing likelihood is computed under the assumption that $LL(\cdot, g)$ follows a Gaussian distribution with a mean determined by the average number of sampled iterations needed to achieve goal $g$. We further assume that long and short thinking times in the goal-recognition experiment correspond respectively to the average number of iterations generated by A-LH for hard and easy goals.

The empirical likelihood draws inspiration from the inverse planning approach introduced by \citet{baker2009action}. We estimate the empirical action and timing likelihoods %through a sampling process, employing real
in the same way as the A-LH likelihoods except that the samples are based
 on human responses collected  during the problem-solving phase instead of simulations from A-LH. For example, the mean and standard deviation for the Gaussian timing likelihood are based on the  human responses provided during the problem-solving phase.

For both the online and empirical likelihoods, we consider variants that incorporate only the action component $LL_A$. These variants are useful for establishing whether timing information is needed to account for our human goal-recognition data.
%the action likelihood
%The terms 'online likelihood with action only' and 'empirical with action only' imply that we consider only the action component $P(\langle a_0, a_1, ..., a_m\rangle|G)$, disregarding timing information.
For all log-likelihood calculations, we add a small value of 0.025 to both options 
%evenly distribute a small value of 0.05 across two options 
to prevent the occurrence of zero probabilities.
\begin{figure*}
\vspace{-3em}
    \centering
    \hspace{-1em}
    \includegraphics[width=\textwidth]{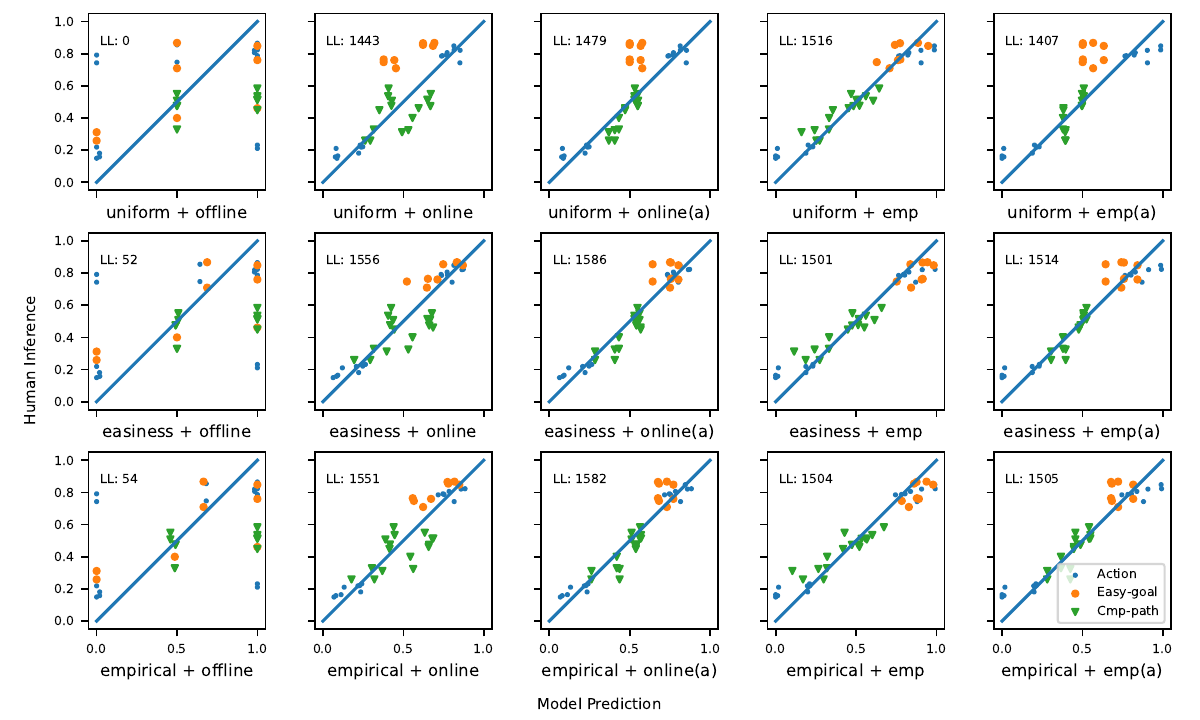}
    \caption{Comparison between model predictions and human inferences. All model labels show the prior followed by the likelihood: for example, \emph{uniform + emp} is the model with uniform Prior and the empirical likelihood. \emph{emp(a)} and \emph{online(a)} are likelihoods that incorporate actions but not timing information. For readability, log likelihoods (higher is better) are shown as offsets relative to the  log likelihood of the \emph{uniform+offline} model.}
    \label{fig:model}
\end{figure*}

\subsubsection{Results and Discussion}
As shown in Fig~\ref{fig:model}, the Easiness prior with online likelihood (actions only) achieves the best overall performance as measured by the log-likelihood assigned to the entire data set. Comparing the rows of Fig~\ref{fig:model} suggests that the contribution of the prior is important but small. In contrast, comparing the columns reveals that changing the likelihood can have a dramatic effect on model performance.  %Nevertheless, it is noteworthy that the selection of the prior has a relatively minimal impact on model fitness while the choice of likelihood has a substantial influence on model performance. 

It is striking that the online likelihoods seems comparable or superior to the empirical likelihoods even though the empirical likelihoods were directly fit to human behavioral data.
%likelihoods lead to similar levels of performance, even though the empirical  of the empirical likelihood, regardless of the incorporation of timing information.
The online likelihoods are based on the A-LH planner, and the strong performance of these likelihoods suggests that the A-LH planner provides a robust and reliable account of human behavior.
%online likelihood (i.e. A-LH) as a planning model to mimic human problem solving, demonstrating its effectiveness in capturing and explaining observed outcomes in the context of goal recognition.
In contrast, the offline likelihood performs substantially worse than the online and empirical likelihoods, suggesting that our  participants implicitly assumed that the actor in the goal-recognition task relied on an online planning strategy.

%that the actor and the inability of this likelihood to capture  which suppo which lacks consideration for human behavior concerning solvability and suboptimal actions, exhibits significant disparities in terms of model fit.

Comparing results for likelihoods with and without timing information suggests that timing information is not needed to account for our behavioral data, and that incorporating this information may slightly impair model performance. Although the online likelihoods with and without timing information yield similar levels of performance, the two show distinct patterns across the three map types. Of the two online likelihoods, the action-only version performs worse across easy-goal maps, but better across the other two map types. This finding suggests that timing information may be beneficial in specific scenarios even though it provided no overall boost in performance across our entire data set.

Although varying the prior does not affect model performance as much as varying the likelihood, it is notable that the Easiness prior model and the empirical prior achieve similar levels of performance. This finding provides additional support for our previous finding (see Fig \ref{fig:model_prior}) that the Easiness model is well-aligned with human judgments.

\section{Related Work}
\citet{ramirez2010probabilistic}, along with subsequent researchers such as \citet{vered2016online} and \citet{masters2019goal}, introduced the Plan Recognition as Planning (PRP) approach that uses planning to estimate the likelihood. We evaluated this approach (referred to as the offline likelihood) as a baseline. This approach assumes agent rationality and focuses exclusively on actions, leaving unaddressed the explicit treatment of unsolvable goals. 

\citet{Zhang_Kemp_Lipovetzky_2023} introduced the Adaptive Lookahead Planner, which was designed to generate human-like response times. We have adapted their planner to incorporate awareness of solvability, and it serves as the online likelihood component in our experiment. While their work explores the impact of timing and how individuals handle timing information within the Sokoban domain, it does not consider the influence of actions and solvability, nor does it provide an explicit and systematic evaluation of the Bayesian approach to goal recognition. \citet{berke2023thinking} have explored the influence of timing information on human understanding of others. Their study, however, is not anchored in the domain of goal recognition, and they rely on a domain-specific algorithm for likelihood estimation.

\citet{baker2017rational} introduced a Bayesian framework for human goal inference and conducted a systematic human experiment demonstrating their model's ability to achieve human-like inference, but did not consider the influence of timing and solvability. They acknowledged the possibility of a non-uniform prior in humans, but did not explore this idea experimentally.

Some recent research has considered non-uniform priors in goal recognition~\cite{gusmao2021inferring, masters2021extended}. These approaches, however, focus mainly on incorporating past information into the prior within the context of  sequential Bayesian updating. We depart from this approach by investigating how domain-independent factors (i.e solvability and easiness) influence human priors.

\section{Conclusion}
In this study we used a Bayesian framework to systematically investigate the influence of actions, timing, and goal solvability on goal recognition. Through an in-depth analysis of human responses in the Sokoban domain, we found that while actions are typically attributed the highest importance, timing and goal solvability also influence goal recognition, particularly in scenarios where actions offer limited information. Leveraging these insights, we developed a goal recognition model that closely aligns with human inferences, surpassing the performance of existing algorithms.

Our work departed from the conventional assumption of a uniform prior, and our results suggest that humans rely on a prior that incorporates factors such as solvability and perceived goal difficulty. We formulated a model of the prior (the Easiness model) that proved successful in accounting for human responses, both before and after any actions had been observed. 
% However, our model comparison results also indicate that, for the purpose of modeling human inference, uniform priors may remain a viable and pragmatic choice. It is essential to emphasize that this observation should not be extended to numerous real-world scenarios. Consider, for instance, a personalized assistant that relies on a uniform prior; in such cases, the assistant's functionality would be severely limited and ineffective.

We extended the Adaptive Lookahead Planner to capture human behavior in the presence of unsolvable goals, and our model comparison suggests that this extended model is useful for estimating the likelihood term in the Bayesian goal recognition framework. This planner, however, departs from human behavior in some respects (e.g.\ by taking more steps before recognizing a goal as unsolvable). This could impact the generalization of our results to real word interactions, and future work should aim to improve it further.

%While our current model has offered valuable insights, further research efforts should focus on reducing the inconsistency between model behavior and human behavior.

Our evaluation of the influence of actions, timing, and solvability suggested that actions (when available) have a dominant influence on people's choices. This finding provides some justification for the standard emphasis on actions within the goal-recognition literature. Nevertheless, our observations also revealed the influence of solvability and timing, particularly in situations where actions are uninformative. Our results seem broadly compatible with an information-seeking approach~\cite{case2016looking} to goal-recognition in which humans focus initially on actions but turn to other factors such as timing and solvability if actions prove uninformative.
%which assumes humans tend to actively gather information to inform their decision-making, as opposed to a traditional Bayesian inference. 
Future work can explore this information-seeking approach in more detail and compare it with the traditional Bayesian approach.
%Exploring the potential for information-seeking to better explain human behavior, particularly in scenarios involving multiple sources of information, represents a promising avenue for future research.

Finally, we conducted a thorough examination of Bayesian inference and the commonly used mirroring approach (i.e. planning for likelihood estimation) discussed in previous work~\cite{baker2009action,baker2017rational,ramirez2010probabilistic,vered2016online}. Our empirical model, which relies on problem-solving data, exhibits a strong alignment with human goal inference. This finding suggests that humans may indeed rely on Bayesian inference and mirroring to carry out goal-recognition. We also introduced a goal recognition model (the model that combines the easiness prior with the online likelihood) that can be implemented independently of human problem-solving data while generating human-like goal inferences. We expect that this model may prove to be useful in a range of downstream applications, including explainable goal recognition \cite{Alshehri_Miller_Vered_2023} and transparent planning, a process focused on selecting actions that effectively convey the actor's intentions to observers \cite{macnally2018action}. Researchers in these domains may be able to leverage this model to advance the development of more interpretable AI behavior.

\balance

\bibliographystyle{ACM-Reference-Format} 
% \bibliography{sample}

%%%%%%%%%%%%%%%%%%%%%%%%%%%%%%%%%%%%%%%%%%%%%%%%%%%%%%%%%%%%%%%%%%%%%%%%

\end{document}